\documentclass[preprint,amsmath,amssymb,aps]{revtex4-1}

\usepackage{graphicx}
\usepackage{dcolumn}
\usepackage{bm}
\usepackage{fancyvrb}

\begin{document}

\title{Zero index electromagnetic materials}

\author{S. A. R. Horsley$^1$, M. Woolley$^2$}
 \affiliation{$^1$Department of Physics and Astronomy, \\
University of Exeter, Stocker Road, Exeter EX4 4QL\\
 $^2$Department of Natural Sciences, \\
University of Exeter, Stocker Road, Exeter EX4 4QL}

\date{\today}

\begin{abstract}
Here we re--examine one of the most basic quantities in optics; the refractive index.  Considering propagation in a plane, we first develop a general formalism for calculating the direction dependent refractive index in a general bi--anisotropic material.  From this we derive the general condition for achieving zero refractive index in a given direction.  We show that when the zero--index direction is complex valued the material supports waves that can propagate in only one sense, e.g. in only a clockwise direction.  We dub such materials complex axis nihility (CAN) media.  Our condition shows that there are an infinite family of both time reversible and time irreversible homogeneous electromagnetic media that support unidirectional propagation for a particular polarization.  We give examples showing that scattering from such media results in the complete exclusion of partial waves with one sign of the angular momentum, and that interfaces between such media generally support one--way interface states.  As well as giving new sets of material parameters, our simple condition reproduces many of the findings derived using topology, such as unidirectional propagation in gyrotropic media, and spin--momentum locking of evanescent waves.
\end{abstract}

\pacs{Valid PACS appear here}
\maketitle
%
%
\section{Introduction}
A ray of light changes direction when passing from one material into another; it refracts.  Refraction was one of the first effects to be captured in a mathematical law, formulated by the Iraqi physicist Ibn Sahl~\cite{smith2015}.  Despite these early origins of what we now know as Snell's law, the index of refraction $n$ wasn't introduced until nearly a thousand years after Sahl~\cite{young1807}.  Since then Maxwell's theory has made it clear that the refractive index is an important quantity, directly determined by the material parameters.  For isotropic materials it is the square root of the product of permittivity $\epsilon$ and permeability $\mu$: $n=\sqrt{\epsilon\mu}$.  Maxwell's equations also tell us that light is an electromagnetic wave, and the index of refraction is the ratio of the wavelength in free space to the wavelength in the material.
\par
 Due to an expansion in available materials~\cite{cai2009,kadic2019}, the concept of refractive index is slightly more subtle in modern physics.  The index can be complex valued, with its imaginary part indicating the degree to which the material absorbs or amplifies the wave, leading to counter intuitive wave effects that are still the subject of active research~\cite{guo2009,horsley2015}.  Its real part can also be either positive or negative~\cite{veselago1967,pendry2000}, with a negative refractive index giving a reversal of the phase velocity, bending the wave in a direction that is impossible with positive index media.  It is also possible to realize materials where the refractive index is very close to zero~\cite{silveirinha2006,alu2007,edwards2008,maas2013}, where the wavelength becomes arbitrarily large and the concept of the ray breaks down.  Close to zero refractive index allows distant points to be electromagnetically close, as is evident in Maxwell's Fish Eye lens~\cite{niven1890}, where infinity is electromagnetically equivalent to a single point~\cite{leonhardt2010}.  Near perfect transmission of electromagnetic waves can be achieved through small zero index channels~\cite{edwards2008}, and unusual boundary conditions can be realised, such as those of an effective magnetic conductor~\cite{liberal2017}.  Such small refractive indices are typically achieved using materials where the permittivity is close to zero, but an effective zero index can also be achieved in waveguides a frequencies close to cut--off~\cite{edwards2008} and layered media close to the edge of a stop--band~\cite{popov2019}.  In general  materials are also anisotropic, meaning that the refractive index depends on the direction of propagation~\cite{born2013}, and the refractive index may be negative or zero for only a small range of propagation directions.  
 \par
 Near zero index materials are the subject of this work.  We begin by considering the problem of finding the material parameters necessary for the index to be zero in some desired direction.  This question is motivated by recent work on one--way propagation in homogeneous media~\cite{davoyan2013,silveirinha2015}, the onset of which can be associated with a zero refractive index in a \emph{complex} direction~\cite{Horsley2017UnidirectionalAxes}, and is also connected to the point in reciprocal space where the Berry curvature is most highly concentrated~\cite{horsley2018b}. Typically such media have been designed using the mathematics of topology, which although fascinating, can in practice involve rather cumbersome calculations.  It is much simpler to calculate a zero in the refractive index than to calculate a Chern number.  Here we look to develop a general understanding of zeros in the refractive index, before using this to design materials that support unidirectional propagation.  We shall find a general condition for such media, before treating some examples and illustrating how the condition reproduces many of the existing findings in the topological photonics literature.
 %
 %
 \section{The $6$--vector form of Maxwell's equations}
 \par
 The electromagnetic properties of a general linear material can be characterised in terms of its constitutive relations.  These relate the displacement field $\boldsymbol{D}$ and magnetic flux density  $\boldsymbol{B}$ to the electric field $\boldsymbol{E}$ and the magnetizing field $\boldsymbol{H}$.  The most general local constitutive relations for \emph{lossless} media take the form
\begin{align}
	\boldsymbol{D}&=\epsilon_0\boldsymbol{\epsilon}\cdot\boldsymbol{E}+\frac{1}{c}\boldsymbol{\xi}\cdot\boldsymbol{H}\nonumber\\
    \boldsymbol{B}&=\mu_0\boldsymbol{\mu}\cdot\boldsymbol{H}+\frac{1}{c}\boldsymbol{\xi}^{\dagger}\cdot\boldsymbol{E}\label{eq:constitutive_relations}.
\end{align}
where $\boldsymbol{\epsilon}$ and $\boldsymbol{\mu}$ are Hermitian tensors and $\boldsymbol{\xi}$ is an arbitrary complex tensor.  Our formalism applies equally well to lossy media, but for simplicity here we restrict ourselves to the lossless case.  A material with a general $\boldsymbol{\epsilon}$, $\boldsymbol{\mu}$ and non--zero tensorial $\boldsymbol{\xi}$, is known as bi--anisotropic~\cite{Mackay2009Electromagneticbianisotropy}.  Perhaps the oldest example of a bi--anisotropic medium is given by the constitutive relations of a moving dielectric~\cite{Landau1984ElectrodynamicsMedia}, where the relativistic transformation of the electromagnetic field naturally mixes the polarization and magnetization.  The effective medium description of a collection of small chiral inclusions is also of this form~\cite{tretyakov1995}.  More recently, it has been recognized that such constitutive relations characterize common metamaterial structures such as arrays of split ring resonators~\cite{pendry1999,Marques2002RoleMetamaterials}.
\par
Now consider the problem of finding the propagation characteristics of waves in homogeneous media described by (\ref{eq:constitutive_relations}).  Given that material parameters are generally a function of frequency, we take fields of a fixed frequency $\omega$, replacing the time derivatives in Maxwell's equations according to the substitution $\partial_t\to-{\rm i}\omega$, giving
\begin{align}
	\boldsymbol{\nabla}\times\boldsymbol{E}&={\rm i}\omega\boldsymbol{B}\nonumber\\
    \boldsymbol{\nabla}\times\boldsymbol{H}&=-{\rm i}\omega\boldsymbol{D}.\label{eq:maxwell}.
\end{align}
For the purposes of calculating the refractive index we write these relations more compactly in terms of a single $6$--vector $F=\left(\boldsymbol{E},\eta_0\boldsymbol{H}\right)^{\rm T}$
\begin{equation}
	\mathcal{D} F=k_0 \chi F\label{eq:6_vector_maxwell}
\end{equation}
where $\mathcal{D}$ and $\chi$ are the Hermitian operators
\begin{equation}
	\mathcal{D}=\left(\begin{matrix}\boldsymbol{0}&{\rm i}\boldsymbol{\nabla}\times\\-{\rm i}\boldsymbol{\nabla}\times&\boldsymbol{0}\end{matrix}\right)
\end{equation}
and
\begin{equation}
    \chi=\left(\begin{matrix}\boldsymbol{\epsilon}&\boldsymbol{\xi}\\\boldsymbol{\xi}^{\dagger}&\boldsymbol{\mu}\end{matrix}\right).\label{eq:chi_def}
\end{equation}
Note that, as discussed in~\cite{barnett2014,horsley2018b,horsley2019,mechelen2019}, Maxwell's equations in the form given in Eq. (\ref{eq:6_vector_maxwell}) have a great deal in common with the Dirac equation used in high energy physics ($\mathcal{D}$ being analogous to the first order operator $-{\rm i}\gamma^{\mu}\partial_{\mu}$ in the Dirac equation).
\par
In an infinite homogeneous medium, the field $F$ can be assumed to have an $\exp({\rm i}k\boldsymbol{n}\cdot\boldsymbol{x})$ dependence, where the wave vector $\boldsymbol{k}=k\boldsymbol{n}$ has magnitude $|k|$ and direction $\boldsymbol{n}$.  In general the magnitude of the wave--vector will depend on the direction of propagation.  Substituting this form of the field in Eq. (\ref{eq:6_vector_maxwell}), we see that the ratio $k_0/k$ can be calculated as an eigenvalue problem
\begin{equation}
	 \chi^{-1} N(\boldsymbol{n}) F=\frac{k_0}{k} F\label{eq:eigenvalue_index}
\end{equation}
where
\begin{equation}
	N(\boldsymbol{n})=\left(\begin{matrix}\boldsymbol{0}&-\boldsymbol{n}\times\\\boldsymbol{n}\times&\boldsymbol{0}\end{matrix}\right).\label{eq:N_operator}
\end{equation}
The $6\times6$ matrix on the left of Eq. (\ref{eq:eigenvalue_index}) has six eigenvalues $k_0/k_{m}$ and eigenvectors $F_{m}$.  Given the symmetry of the matrix $N(\boldsymbol{n})$ we can see from Eq. (\ref{eq:eigenvalue_index}) that for real eigenvalues we have both $N F_{n}=(k_0/k_n) \chi F_n$, and $F_{m}^{\dagger}N=(k_0/k_m)F_{m}^{\dagger}\chi$.  Therefore when the eigenvalues are real, the eigenvectors $F_n$ are orthogonal with respect to the inner product, allowing us to write
\begin{equation}
    F_{n}^{\dagger}\chi F_{m}=\pm\delta_{nm}\qquad(n\neq m).\label{eq:normalization}
\end{equation}
where the sign of the right hand side of (\ref{eq:normalization}) is not necessarily positive when $n=m$ because e.g. $\chi$ could be negative definite.  Two of the eigenvalues of Eq. (\ref{eq:eigenvalue_index}) are zero, corresponding to eigenvectors $(\boldsymbol{n},\boldsymbol{0})^{\rm T}$ and $(\boldsymbol{0},\boldsymbol{n})^{\rm T}$.  These are the electrostatic and magnetostatic modes, where either $k_0=0$, or $k\to\infty$.  The remaining four eigenvalues come as two pairs of opposite sign.  These correspond to electromagnetic waves, with a negative sign of $k$ being equivalent to a reversal of the propagation direction $\boldsymbol{n}\to-\boldsymbol{n}$, as is clear from the defining equation, Eq. (\ref{eq:eigenvalue_index}).  These four eigenvalues are the two electromagnetic polarizations, propagating in either the $+\boldsymbol{n}$ or $-\boldsymbol{n}$ direction.
\par
The magnitude of the refractive index for each of the modes is given by the magnitude of the inverse of the eigenvalues $k_0/k_m$.  Determining the \emph{sign} of the index requires a bit more thought, and turns out to be related to the indeterminate sign of the normalization in Eq. (\ref{eq:normalization}).  The sign of the index is determined by whether the time averaged Poynting vector $\boldsymbol{S}={\rm Re}[\boldsymbol{E}\times\boldsymbol{H}^{\star}]/2$ is parallel or anti--parallel to the wave--vector $\boldsymbol{k}=k_m\boldsymbol{n}$, i.e. the sign of the index is the sign of
\begin{equation}
    \boldsymbol{k}\cdot\boldsymbol{S}=k_{m}\boldsymbol{n}\cdot\frac{1}{2}{\rm Re}\left[\boldsymbol{E}\times\boldsymbol{H}^{\star}\right]=\frac{1}{4\eta_0}k_{m}F^{\dagger}_{m}\,N(\boldsymbol{n})\,F_{m}
\end{equation}
or, using Eq. (\ref{eq:eigenvalue_index})
\begin{equation}
    {\rm sign}\left[\boldsymbol{k}\cdot\boldsymbol{S}\right]={\rm sign}\left[F^{\dagger}_{m}\,\chi\,F_{m}\right]
\end{equation}
The sign of the refractive index for the eigenmode $F_{m}$ is thus determined by the sign of the inner product $F_{m}^{\dagger}\,\chi\,F_{m}$. Media where every mode propagates with a negative index are thus characterized as having a negative definite material tensor $\chi$.  This recovers the standard result that an isotropic magnetodielectric with
\begin{equation}
    \chi=\chi_{\rm neg}=\left(\begin{matrix}-|\epsilon|\boldsymbol{1}&\boldsymbol{0}\\\boldsymbol{0}&-|\mu|\boldsymbol{1}\end{matrix}\right)
\end{equation}
exhibits a negative refractive index.  Interestingly, it also shows that there are an infinite family of bi--anisotropic materials where the refractive index is negative: for any $6\times6$ matrix $A$ we can write a susceptibility with negative index as $\bar{\chi}_{\rm neg}=-A^{\dagger}A$.  
\par
This concludes the introduction to our formalism.  We note that there is some similarity to the $3\times 3$ formalism of Berry and Dennis used in~\cite{berry2003,berry2005}, although in that work the concern was with singular points where e.g. the refractive indices of different polarizations become degenerate.
%
%
\begin{figure}[h!]
    \centering
    \includegraphics[width=\textwidth]{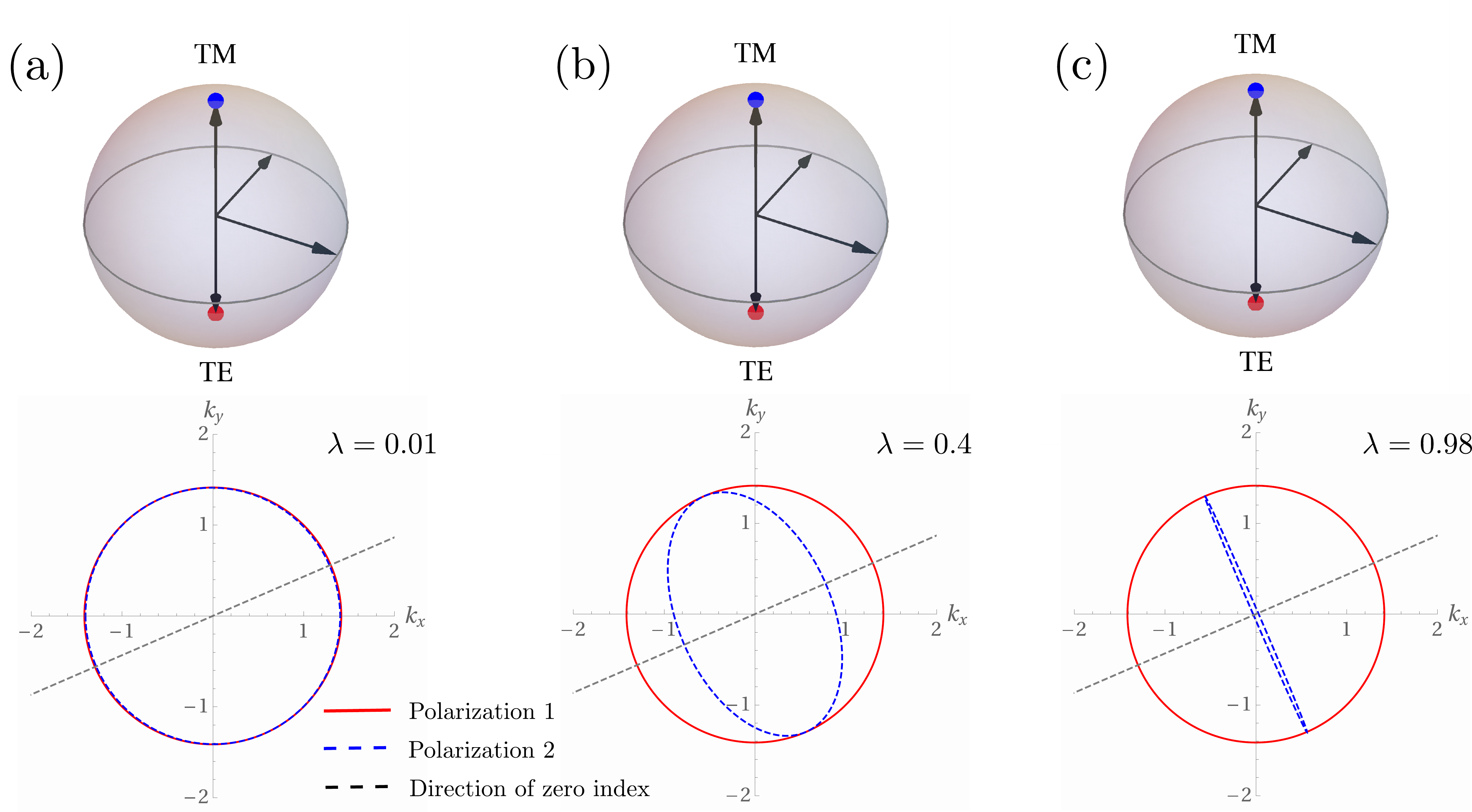}
    \caption{Dispersion relation (lower panels) and polarization (upper panels) as a function of propagation angle.  Starting from a material with $\boldsymbol{\epsilon}=2 1_{3}$, $\boldsymbol{\mu}=1_{3}$ and $\boldsymbol{\xi}=0$, we form the projected material tensor $\chi$ from Eq. (\ref{eq:chi_proj}) with $\psi=0$; $\sigma=\pi/2$ (arbitrary); and $\boldsymbol{m}=0.911\,\hat{\boldsymbol{x}}+0.397\,\hat{\boldsymbol{y}}$.  Computing the dispersion relation from Eq. (\ref{eq:eigenvalue_index}) we see that as $\lambda$ approaches unity the refractive index of waves propagating in the direction indicated by the dashed black line approaches zero.  As indicated by Eq. (\ref{eq:pol1}) and shown in the upper panels the eigenpolarizations (here visualized on the Bloch sphere) have either electric or magnetic field pointing out of the plane.}
    \label{fig:dispersion_example}
\end{figure}
%
%
\section{Anisotropic zero index media}
%
%
\begin{figure}[h!]
    \centering
    \includegraphics[width=\textwidth]{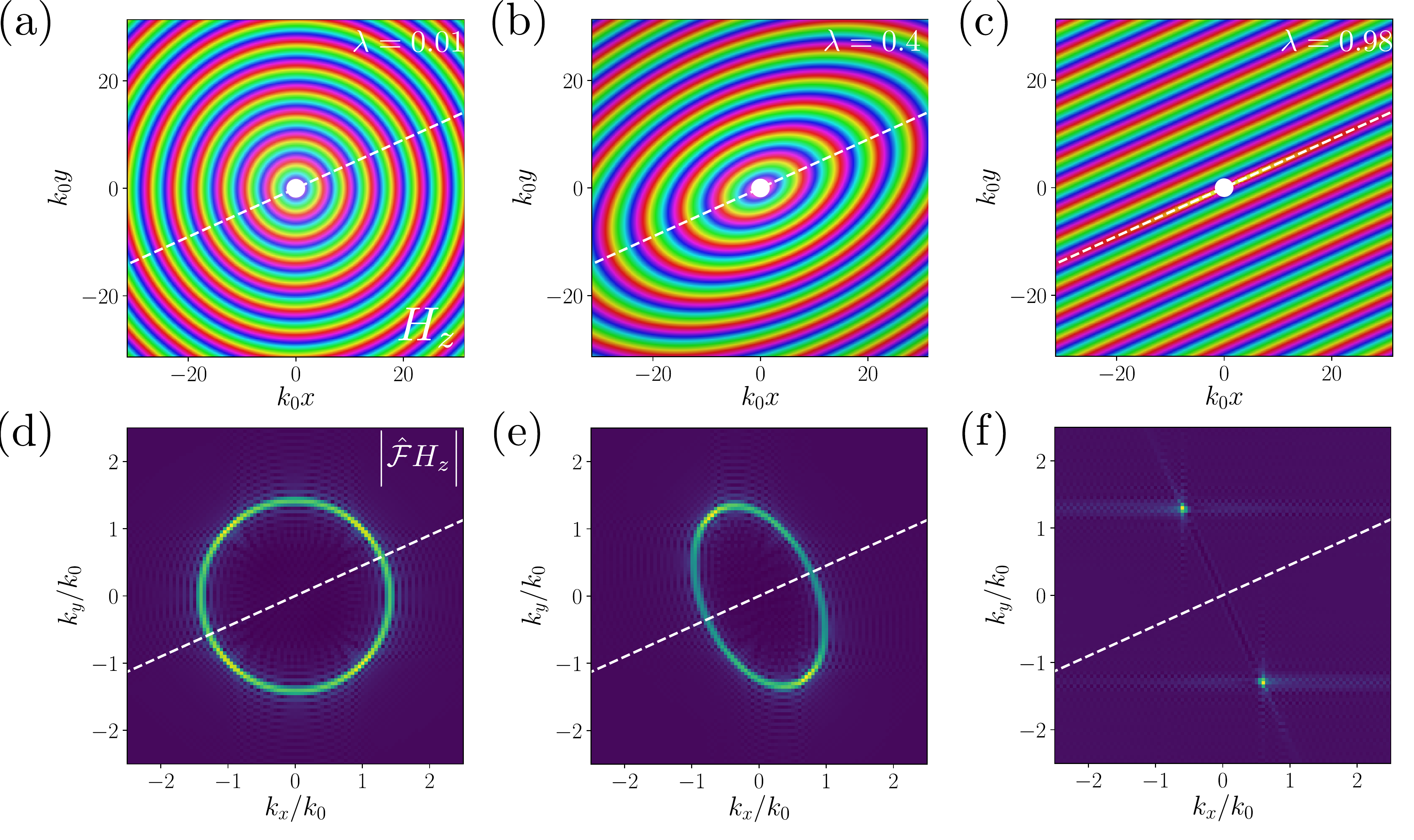}
    \caption{Full wave simulation (using COMSOL multiphysics~\cite{comsol}) of the radiation from an out of plane magnetic current (central white dot) in a homogeneous medium.  The material parameters are those used in Fig.~\ref{fig:dispersion_example}.  Panels (a--c) show the stretching of the out of plane magnetic field along the zero index axis (white dashed line). Panels (d--f) show the Fourier magnitudes (computed using the SciPy~\cite{scipy} FFTPACK library) of the fields given in the upper panels, demonstrating that the dispersion relation is that predicted in Fig.~\ref{fig:dispersion_example}.  Note that the total simulation area is not sufficient to resolve the transverse dimension of the dispersion relation in panel (f). }
    \label{fig:zero_index_point_source}
\end{figure}
\par

%
%
\begin{figure}[h!]
    \centering
    \includegraphics[width=\textwidth]{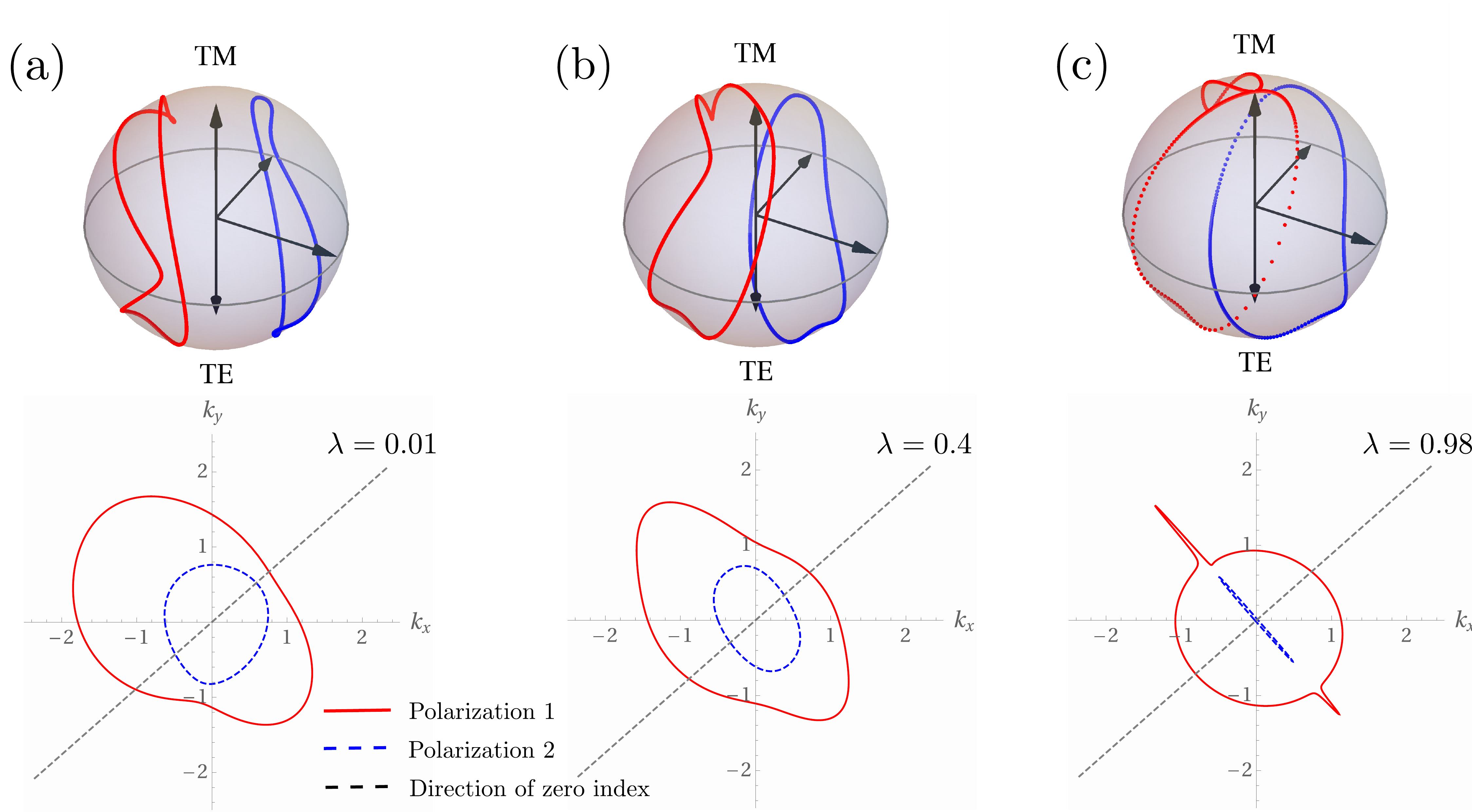}
    \caption{As in Fig.~\ref{fig:dispersion_example}, but where $\chi_{\rm i}$ is the bi--anisotropic material with the randomly generated $6\times6$ Hermitian matrix $\chi_{\rm i}$ given in appendix~\ref{ap:random},  (\ref{eq:chi0_random}).  The projected material matrix $\chi$ is then formed from Eq. (\ref{eq:chi_proj}), with $\psi=\pi/2$; $\sigma=\pi/2$ (arbitrary); and $\boldsymbol{m}=0.750\,\hat{\boldsymbol{x}}+0.661\,\hat{\boldsymbol{y}}$.  In contrast to Fig.~\ref{fig:dispersion_example}, the polarization state (upper panels) changes with angle, and here we plot the state for 628 evenly spaced angles.  In panel (c) it is clear that polarization 1 becomes concentrated around the state given by Eq. (\ref{eq:pol1}), changing very rapidly as the propagation angle becomes close to orthogonal to the direction of zero index, $\boldsymbol{m}$.}
    \label{fig:random_example}
\end{figure}
Now let's consider the problem of finding a zero in the refractive index.  We assume that propagation is in the $x$--$y$ plane, so that the field is determined by the out of plane electric $E_z$ and magnetic $H_z$ field components.  A zero refractive index in a particular direction, $\boldsymbol{m}=m_x\hat{\boldsymbol{x}}+m_y\hat{\boldsymbol{y}}$, implies an infinite stretching of the wavelength for propagation along that axis.  We can calculate the gradient of a linear combination of $E_z$ and $H_z$ along the $\boldsymbol{m}$ axis, through taking the inner product of Maxwell's equations, Eq. (\ref{eq:6_vector_maxwell}) with the following normalized vector
\begin{equation}
    V(\boldsymbol{m},\sigma,\psi)={\rm i}\left(\begin{matrix}\cos(\psi)\boldsymbol{m}\times\hat{\boldsymbol{z}}\\-\sin(\psi){\rm e}^{{\rm i}\sigma}\boldsymbol{m}\times\hat{\boldsymbol{z}}\end{matrix}\right)\label{eq:V_in_plane}
\end{equation}
This gives
\begin{equation}
    V^{\dagger}\mathcal{D}F=\boldsymbol{m}\cdot\boldsymbol{\nabla}\left[\sin(\psi){\rm e}^{-{\rm i}\sigma}E_z+\cos(\psi)\eta_0 H_z\right]=k_0\,V^{\dagger}\chi F=0.\label{eq:grad_m}
\end{equation}
We have assumed that the index is zero in the $\boldsymbol{m}$ direction, so that the field is uniform along that axis.  We can ensure this if $V$ is a zero eigenvector of the Hermitian constitutive tensor $\chi$,
\begin{equation}
    V^{\dagger}\chi=\chi V=0.\label{eq:zero_index_condition}
\end{equation}
Therefore a zero in the refractive index is associated with a zero eigenvalue of the material tensor $\chi$, with the nullspace vector $V$ lying in the plane of propagation, as written in Eq. (\ref{eq:V_in_plane}).  We can thus parameterize the refractive index in the $\boldsymbol{m}$ direction through introducing the projection operator
\begin{equation}
    P(\lambda,\boldsymbol{m},\sigma,\psi)=1_{6}-\lambda V\otimes V^{\dagger}\label{eq:projection}
\end{equation}
Taking any initial material tensor $\chi_{\rm i}$ we can use the projection operator (\ref{eq:projection}) to define a new tensor
\begin{equation}
    \chi(\lambda,\boldsymbol{m},\sigma,\psi)=P\,\chi_{\rm i}\, P\label{eq:chi_proj}
\end{equation}
which equals $\chi_{\rm i}$ when $\lambda=0$, and has zero index in the $\boldsymbol{m}$ direction when $\lambda=1$.  The index is however only zero for one of the two polarizations.  From Eq. (\ref{eq:grad_m}), we can see that for $\lambda=1$, the polarization where
\begin{align}
    E_{z}&={\rm e}^{{\rm i}\sigma}\cos(\psi)\Phi\nonumber\\
    \eta_0 H_z&=-\sin(\psi)\Phi\label{eq:pol1}
\end{align}
does not necessarily have zero index, because the left hand side of (\ref{eq:grad_m}) is identically zero before the gradient is taken ($\Phi$ is an arbitrary complex number).  It is the polarization orthogonal to (\ref{eq:pol1}) with respect to the inner product (\ref{eq:normalization}) that has zero index when $\lambda=1$.  In Figures~\ref{fig:dispersion_example} and~\ref{fig:random_example} we illustrate the above results with two numerical examples.  In the first example we start with an isotropic dielectric ($\epsilon=2$ and $\mu=1$), projecting out a prespecified direction and polarization angle according to (\ref{eq:chi_proj}), until the index is zero for propagation at the arbitrarily chosen angle of $23$ degrees to the $x$ axis.  As $\lambda$ approaches $1$ this results in an increasingly anisotropic material where the dispersion circle is squashed into a infinitely thin ellipse.  Figure~\ref{fig:zero_index_point_source} demonstrates the agreement between this prediction and a full wave simulation.  The same effect (this time for propagation at $41$ degrees to the $x$ axis) is demonstrated in Fig.~\ref{fig:random_example}, but there starting from an arbitrary material tensor $\chi_{\rm i}$, generated using a random number generator, illustrating the general applicability of our zero--index condition.
%
%
\section{Zero index in a complex direction\label{sec:complex_zero_index}}
%
%
\begin{figure}[h!]
    \centering
    \includegraphics[width=0.8\textwidth]{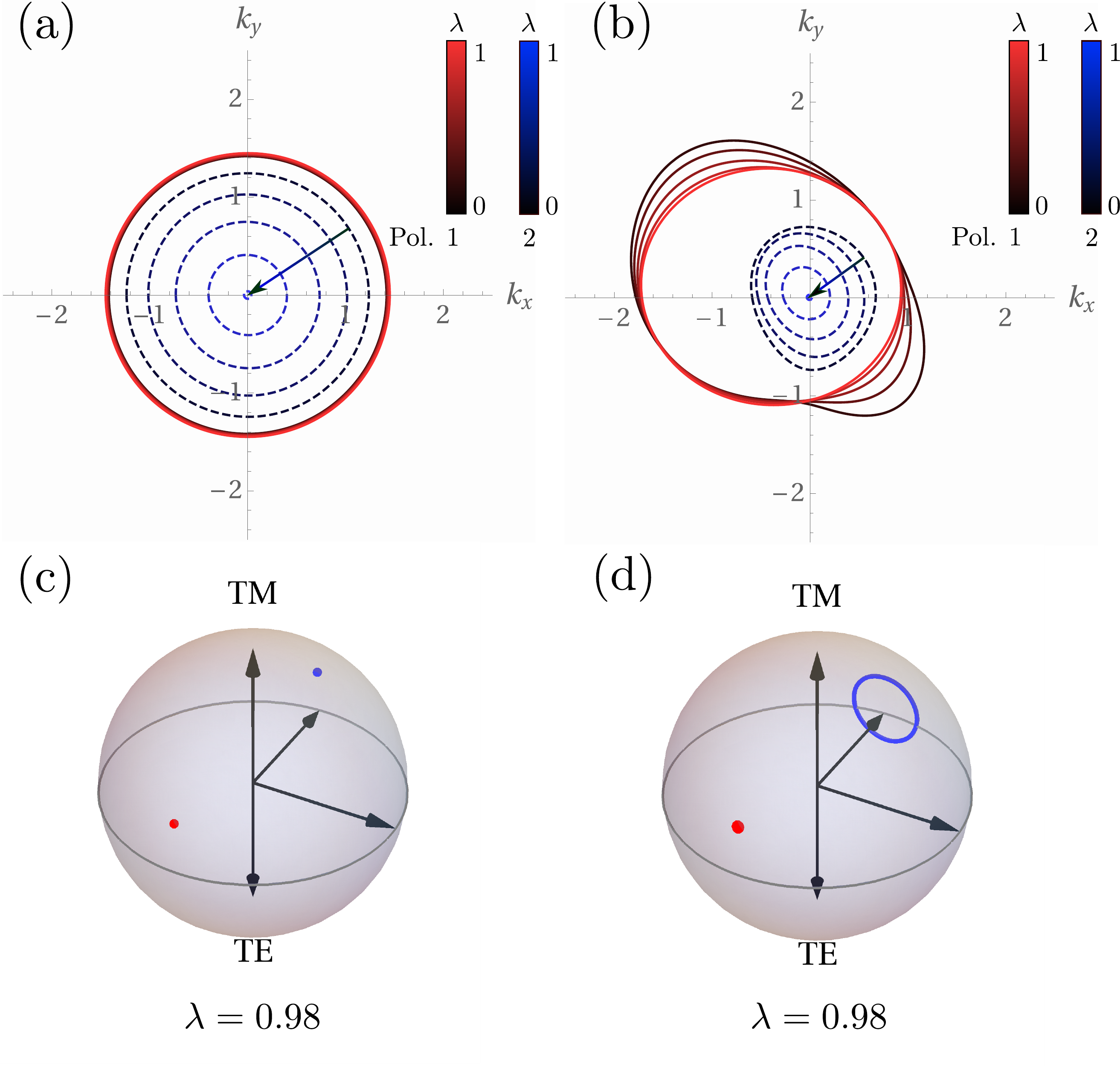}
    \caption{As in Figs.~\ref{fig:dispersion_example} and \ref{fig:random_example}, but approaching zero index in the arbitrarily chosen complex direction, $\boldsymbol{m}=(0.656-0.286{\rm i})\hat{\boldsymbol{x}}+(0.296+0.633{\rm i})\hat{\boldsymbol{y}}$.  The remaining parameters are $\psi=3\pi/10$ and $\sigma=\pi/2$.  In panels (a) and (c) the initial material ($\lambda=0$) is that of Fig.~\ref{fig:dispersion_example}, and in panels (b) and (d) it is the randomly generated material of Fig.~\ref{fig:random_example}.  Note that panel (d) shows the polarization of the zero index mode may not become fixed as $\lambda\to1$.}\label{fig:complex-zero-index}
\end{figure}
\par
Now take a different case of our zero index condition Eq. (\ref{eq:grad_m}).  Let the zero index direction $\boldsymbol{m}$ be \emph{complex} valued.  For brevity we'll refer to such materials as complex axis nihility (CAN) media.  As a concrete example, consider the vector $\boldsymbol{m}=(\hat{\boldsymbol{x}}+{\rm i}\hat{\boldsymbol{y}})/\sqrt{2}$, which over time traces out a circle in the $x$--$y$ plane.  This special case was recently associated with uni--directional wave propagation and photonic edge states in gyrotropic media~\cite{Horsley2017UnidirectionalAxes}.  For this particular form of $\boldsymbol{m}$, the out of plane electric and magnetic fields satisfy the following form of Eq. (\ref{eq:grad_m})
\begin{equation}
    \left(\frac{\partial}{\partial x}+{\rm i}\frac{\partial}{\partial y}\right)\left[\sin(\psi){\rm e}^{-{\rm i}\sigma}E_{z}+\cos(\psi)\eta_0 H_z\right]=2\frac{\partial}{\partial\mathcal{Z}^{\star}}\left[\sin(\psi){\rm e}^{-{\rm i}\sigma}E_{z}+\cos(\psi)\eta_0 H_z\right]=0\label{eq:analyticity}
\end{equation}
where $\mathcal{Z}=x+{\rm i}y$.  Equation (\ref{eq:analyticity}) is equivalent to the statement that either the polarization is of the form (\ref{eq:pol1}), or the out of plane field is an analytic function of position satisfying the Cauchy--Riemann conditions.  This has a rather different interpretation to the previous cases where $\boldsymbol{m}$ was real.  In this case we do not squash the dispersion circle into a line, or stretch the field out to uniformity along one axis.  Instead the out of plane field is allowed to propagate in only one sense around the origin.  This is evident when we write the Taylor expansion of the field in polar coordinates $\mathcal{Z}=r\,{\rm e}^{{\rm i}\theta}$
\begin{align}
    \sin(\psi){\rm e}^{-{\rm i}\sigma}E_{z}+\cos(\psi)\eta_0 H_z&=\sum_{n=0}^{\infty}c_{n}\mathcal{Z}^{n}\nonumber\\
    &=\sum_{n=0}^{\infty}c_{n}\,r^{n}\,{\rm e}^{{\rm i}n\theta}
\end{align}
Because the series contains only positive $n$ (a restriction that comes from demanding no singularities in the field, and requires the medium to be simply connected), the series expansion contains only terms with phases that wind anti--clockwise around the origin.  Therefore a simply--connected material with zero refractive index in a complex direction $\boldsymbol{m}=\boldsymbol{m}'+{\rm i}\boldsymbol{m}''$ will support waves that have a fixed sense of rotation around every point in the medium.  This is a rather straightforward way to characterize materials exhibiting unidirectional propagation.  For media of this type, the dispersion surface does not form the flattened ellipse as $\lambda\to1$ (shown in Figs.~\ref{fig:dispersion_example} and \ref{fig:random_example}), instead it closes to a point, an effect which is illustrated in Fig.~\ref{fig:complex-zero-index}, and can be understood in terms of the zero index condition (\ref{eq:grad_m}).  For a plane wave, condition (\ref{eq:grad_m}) simplifies to
\begin{equation}
    \left(\boldsymbol{m}'+{\rm i}\boldsymbol{m}''\right)\cdot\boldsymbol{k}=0
\end{equation}
which---unless $\boldsymbol{m}'$ and $\boldsymbol{m}''$ are parallel---cannot be satisfied for any real valued $\boldsymbol{k}$, except $\boldsymbol{k}=0$.  This is similar to the gyrotropic case of the optical Dirac equation explored in~\cite{horsley2018b}, where unidirectional propagation occurs within the region of parameter space where the effective energy is less than the effective mass, and propagation is forbidden.  At the boundaries of this region of parameter space the dispersion relation is $\boldsymbol{k}=0$, and the wave becomes an analytic function of position as in Eq. (\ref{eq:analyticity}).  One can also relate the above findings to existing work on the effective zero index condition that occurs within periodic media in the vicinity of the Dirac point~\cite{huang2011}.
\par
We'll now derive some physical consequences of having zero index in a complex direction.  To be concrete we'll consider two specific cases where (\ref{eq:grad_m}) holds.  The simplest case is a gyrotropic medium where e.g. $\boldsymbol{\epsilon}=\boldsymbol{1}_{3}\pm{\rm i}\lambda\hat{\boldsymbol{z}}\times$, $\boldsymbol{\mu}=\boldsymbol{1}_{3}$ and $\boldsymbol{\xi}=\boldsymbol{0}$.  However, there is already a great deal of literature that explores unidirectional propagation in gyrotropic media~\cite{davoyan2013,silveirinha2015,silveirinha2016,Horsley2017UnidirectionalAxes,horsley2018b}.  Instead we'll take two experimentally accessible bi--anisotropic materials that illustrate the general applicability of Eq. (\ref{eq:grad_m}).
%
%
\subsection{Scattering from a CAN cylinder}
\par
A simple example of a CAN medium is one where the sum of the out of plane field components $E_{z}+\eta_0 H_{z}$ exhibits unidirectional propagation,
and as in Eq. (\ref{eq:analyticity}) we'll take the zero index direction as $\boldsymbol{m}=(\hat{\boldsymbol{x}}+{\rm i}\hat{\boldsymbol{y}})/\sqrt{2}$.  Expanding out our zero index condition (\ref{eq:grad_m}) we find
\begin{equation}
    \sqrt{2}\frac{\partial}{\partial\mathcal{Z}^{\star}}\left(E_{z}+\eta_0 H_z\right)=k_0\left[\boldsymbol{m}\cdot(\boldsymbol{\epsilon}-\boldsymbol{\xi}^{\dagger})\cdot\boldsymbol{E}-\boldsymbol{m}\cdot\left(\boldsymbol{\mu}-\boldsymbol{\xi}\right)\cdot\eta_0\boldsymbol{H}\right]=0\label{eq:example_1_analyticity}
\end{equation}
which shows that our zero index condition is equivalent to
\begin{align}
    \boldsymbol{m}\cdot\left(\boldsymbol{\epsilon}-\boldsymbol{\xi}^{\dagger}\right)&=0\nonumber\\
    \boldsymbol{m}\cdot\left(\boldsymbol{\mu}-\boldsymbol{\xi}\right)&=0\label{eq:zero_index_condition_2}.
\end{align}
There are many possible ways to fulfill these conditions, but we define the particular set of material parameters
\begin{equation}
    \boldsymbol{\epsilon}=\boldsymbol{1}_{3},\;\boldsymbol{\mu}=\boldsymbol{1}_{3},\;\boldsymbol{\xi}=\lambda\left(\begin{matrix}0&{\rm i}&0\\-{\rm i}&0&0\\0&0&0\end{matrix}\right)\label{eq:zero_index_material}.
\end{equation}
As in Sec.~\ref{sec:complex_zero_index}, in the limit $\lambda\to1$ this material satisfies the zero index condition (\ref{eq:zero_index_condition_2}).  However, note that we did not form (\ref{eq:zero_index_material}) through taking the projection (\ref{eq:projection}).  Unlike a gyrotropic material, this material is time reversible: taking $\boldsymbol{H}\to-\boldsymbol{H}$ and ${\rm i}\to-{\rm i}$ leaves the constitutive relations (\ref{eq:constitutive_relations}) unchanged.  In fact, the bi--anisotropic response given in (\ref{eq:zero_index_material}) is known at microwave frequencies to be that of an array of small Omega shaped wire particles~\cite{saadoun1992,tretyakov1993} (our particular case is rather special because it also requires $\boldsymbol{\epsilon}=\boldsymbol{\mu}$).  Time reversibility may seem completely at odds with what we're trying to do: how is it possible that there is unidirectional propagation in a time reversible material?  If we reverse time, the wave propagates in the reverse direction, while the material remains unchanged!  We achieve this is the same manner as Ref.~\cite{liu2015}.  The time irreversiblity comes from the polarization basis, which upon reversal undergoes the transformation $E_{z}+\eta_0 H_{z}\to E_{z}-\eta_0 H_z$.  As a consequence we must therefore have the combinations of the out of plane fields, $E_{z}+\eta_0 H_{z}$ and $E_{z}-\eta_0 H_{z}$ propagating in opposite senses; one becoming an analytic function of $\mathcal{Z}$, the other of $\mathcal{Z}^{\star}$.
%
%
\begin{figure}[h!]
    \centering
    \includegraphics[width=0.8\textwidth]{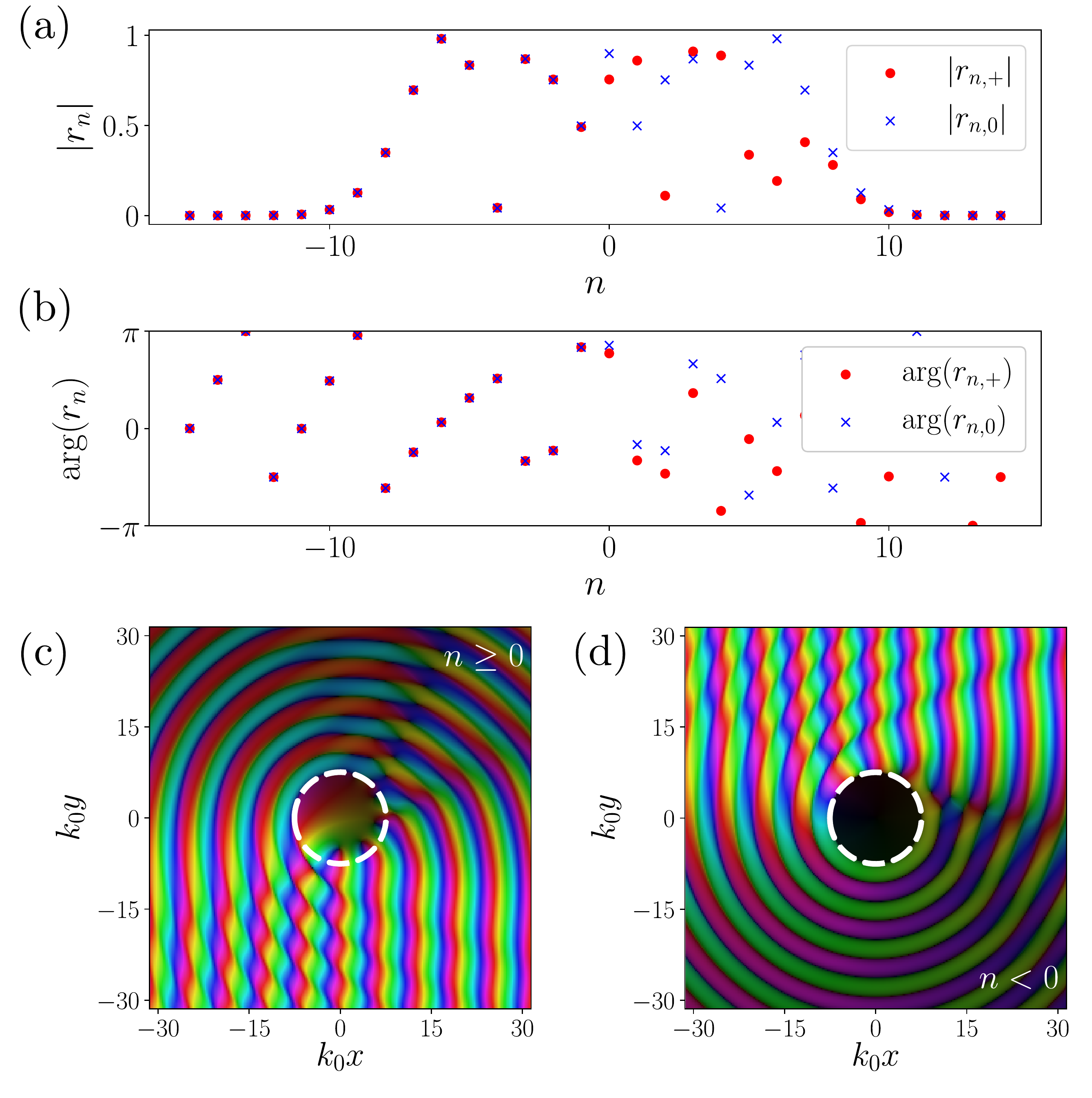}
    \caption{Scattering of the $\Psi_{+}$ polarization from the bi--anisotropic CAN medium (\ref{eq:zero_index_material}) (equivalent to an Omega medium~\cite{saadoun1992}), with $\lambda=0.999$.  The radius of cylinder is $k_0 a=7.54$, and $r_{n,0}$ refers to the partial wave scattering amplitude obtained when the wave satisfies $\Psi_{+}=0$ on the surface of the cylinder. Panels (a) and (b) show that the magnitude and phase of the scattered partial wave amplitudes $r_{n,+}$ are equal to $r_{n,0}$ for $n>0$.  Panels (c) and (d) plot the positive and negative angular momentum parts of the field $\Psi_{+}$ (evaluated as partial sums of (\ref{eq:cylinder_field})), showing that the negative angular momenta are completely excluded from the cylinder (field amplitude indicated as saturation and phase as colour $(0,\pi/2,\pi,3\pi/2)\to({\rm red},{\rm green},{\rm cyan},{\rm purple})$).}
    \label{fig:scattering}
\end{figure}
\par
Suppose we scatter radiation from a cylinder of radius $a$ composed of the bi--anisotropic material defined in (\ref{eq:zero_index_material}).  Within the cylinder, Maxwell's equations can be written in the form
\begin{equation}
    \boldsymbol{\nabla}\Psi_{\pm}\times\hat{\boldsymbol{z}}={\rm i}k_0\left(\boldsymbol{1}_{2}\pm{\rm i}\lambda\hat{\boldsymbol{z}}\times\right)\cdot\boldsymbol{\Phi}_{\mp}\label{eq:out_of_plane_omega}
\end{equation}
and
\begin{equation}
    \boldsymbol{\nabla}_{\parallel}\times\boldsymbol{\Phi}_{\pm}=-{\rm i}k_0 \Psi_{\mp}\label{eq:in_plane_omega}
\end{equation}
where $\Psi_{\pm}=E_{z}\pm\eta_0 H_z$, and $\boldsymbol{\Phi}_{\pm}=\eta_0\boldsymbol{H}_{\parallel}\pm\boldsymbol{E}_{\parallel}$, with `$\parallel$' indicating field components in the plane of propagation.  Taking the inner product of Eq. (\ref{eq:out_of_plane_omega}) with the matrix $\boldsymbol{1}_{2}\mp{\rm i}\lambda\hat{\boldsymbol{z}}\times$, 
gives us the in--plane field components $\boldsymbol{\Phi}_{\pm}$ in terms of derivatives of the out of plane ones $\Psi_{\pm}$
\begin{equation}
    \boldsymbol{\nabla}\Psi_{\pm}\times\hat{\boldsymbol{z}}\mp{\rm i}\lambda\boldsymbol{\nabla}\Psi_{\pm}={\rm i}k_0\left(1-\lambda^{2}\right)\boldsymbol{\Phi}_{\mp}\label{eq:in-plane-fields}
\end{equation}
taking the curl of both sides and then applying (\ref{eq:in_plane_omega}) yields the equation for the out of plane field in the cylinder
\begin{equation}
    \boldsymbol{\nabla}^{2}\Psi_{\pm}+k_0^{2}\left(1-\lambda^{2}\right)\Psi_{\pm}=0\label{eq:CAN_helmholtz}
\end{equation}
which is simply the Helmholtz equation for a scalar wave in a medium with refractive index $n=\sqrt{1-\lambda^{2}}$.  Inside and outside of the cylinder we can therefore use cylindrical coordinates $(r,\theta)$, and expand the wave as a sum of Bessel functions
\begin{equation}
    \Psi_{\pm}(r,\theta)=\sum_{n=-\infty}^{\infty}{\rm e}^{{\rm i}n\theta}
    \begin{cases}
    {\rm i}^{n}J_{n}(k_0 r)+r_{n,\pm} \mathcal{H}_n^{(0)}(k_0 r)&\qquad r>a\\
    \left[{\rm i}^{n}J_{n}(k_0 a)+r_{n,\pm} \mathcal{H}_n^{(0)}(k_0 a)\right] \frac{J_{n}\left(kr\right)}{J_{n}\left(ka\right)}&\qquad r<a
    \end{cases}\label{eq:cylinder_field}
\end{equation}
where $k=\sqrt{1-\lambda^{2}}k_0$, and $J_n$ and $\mathcal{H}_n^{(0)}$ are Bessel and Hankel functions of the first kind, respectively.  The particular combination of terms given in Eq.~(\ref{eq:cylinder_field}) assumes for $r>a$ an incident wave, $\exp({\rm i}k_0 x)=\sum_{n}{\rm i}^{n}J_n(k_0 r)\exp({\rm i}n\theta)$~\cite{dlmf}, plus an outgoing wave with partial wave amplitudes $r_n$.  Inside the cylinder where $r<a$ we take the Bessel functions that are non--singular at the origin, $J_n$, with amplitudes such that $\Psi_{\pm}$ is continuous at $r=a$.  The scattering amplitudes $r_{n,\pm}$ are determined by the remaining Maxwell boundary condition: the continuity of the $\hat{\boldsymbol{\theta}}\cdot\boldsymbol{\Phi}_{\pm}$.  Applying this condition to Eq. (\ref{eq:in-plane-fields}) and then using our partial wave expansion (\ref{eq:cylinder_field}) we find the partial wave amplitudes are equal to
\begin{equation}
    r_{n,\pm}=-{\rm i}^{n}\frac{J_n'(k_0 a)-J_n(k_0 a)\Gamma_{n,\pm}}{\mathcal{H}_{n}'^{(0)}(k_0 a)-\mathcal{H}_{n}^{(0)}(k_0 a)\Gamma_{n,\pm}}\label{eq:partial_wave_amplitudes}
\end{equation}
where
\begin{equation}
    \Gamma_{n,\pm}=\frac{1}{1-\lambda^{2}}\left(\frac{k}{k_0}\frac{J_n'(k a)}{J_n(ka)}\mp\frac{n\lambda}{k_0 a}\right)\label{eq:gamma_def}.
\end{equation}
In the limit where $\Gamma_{n,\pm}\to \infty$, the partial wave amplitudes are the same as if the field component $\Psi_{\pm}$ was set to zero on the surface of the cylinder, i.e. the boundary conditions $E_{z}=\mp\eta_0 H_z$.  A cursory examination of Eq. (\ref{eq:gamma_def}) suggests this occurs in our zero index limit $\lambda\to1$, due to the divergence of the prefactor $1/(1-\lambda^2)$.  However, a more careful investigation of the limit shows this is only true for one sign of angular momentum $n$.  Taking $\lambda=1-\eta$ ($\eta\ll1$) and expanding Eq. (\ref{eq:gamma_def}) to leading order gives
\begin{equation}
    \Gamma_{n,+}\sim\begin{cases}
    \frac{|n|}{2k_0a}-\frac{k_0 a}{2(|n|+1)}&\qquad n\geq0\\
    \frac{|n|}{\eta k_0 a}&\qquad n<0
    \end{cases}\label{eq:g1}
\end{equation}
and
\begin{equation}
    \Gamma_{n,-}\sim\begin{cases}
    \frac{|n|}{\eta k_0 a}&\qquad n>0\\
    \frac{|n|}{2k_0a}-\frac{k_0 a}{2(|n|+1)}&\qquad n\leq0
    \end{cases}\label{eq:g2}
\end{equation}
where we used the series expansion of $J_n(x)$ up to second order in $x$~\cite{dlmf}.  As we let $\eta\to 0$, Eqns. (\ref{eq:g1}--\ref{eq:g2}) shows that there is a divergence for only one sign of $n$, and therefore the partial waves with one sign of the angular momentum are set to zero on the surface of the cylinder.  We expect the polarization $\Psi_{+}$ to become an analytic function of $\mathcal{Z}=x+{\rm i}y=r\,{\rm e}^{{\rm i}\theta}$ within the cylinder, and accordingly the waves with a clockwise winding phase are completely reflected, i.e. $\Gamma_{n<0,+}\to\infty$.  The reverse holds for the anticlockwise winding waves of polarization $\Psi_{-}$.  Fig.~\ref{fig:scattering} illustrates this phenomenon, plotting for the $\Psi_{+}$ polarization, both the scattered partial wave amplitudes (\ref{eq:partial_wave_amplitudes}) and the positive/negative angular momentum parts of the field (\ref{eq:cylinder_field}).  This demonstrates that a CAN medium is one which forbids the propagation of waves that rotate in one sense; in this particular case the $n<0$ waves of the $\Psi_{+}$ polarization are completely excluded from the volume of the scattering material.
%
%
\subsection{Unidirectional edge states and evanescent waves}
\par
As a second example of CAN media we demand that the field $\Psi'_{+}=E_{z}+{\rm i}\alpha\eta_0 H_{z}$ propagates with zero index, again in the $\boldsymbol{m}=\hat{\boldsymbol{x}}+{\rm i}\hat{\boldsymbol{y}}$ direction.  Expanding out condition (\ref{eq:grad_m}) we find, similar to the previous section
\begin{align}
    \boldsymbol{m}\cdot\left(\boldsymbol{\epsilon}+{\rm i}\alpha^{-1}\boldsymbol{\xi}^{\dagger}\right)&=0\nonumber\\
    \boldsymbol{m}\cdot\left(\boldsymbol{\mu}-{\rm i}\alpha\boldsymbol{\xi}\right)&=0\label{eq:condition_example2}
\end{align}
a condition which can again be fulfilled in many ways.  We make the following choice of material parameters, which is only a factor of ${\rm i}$ different from (\ref{eq:zero_index_material})
\begin{align}
    \boldsymbol{\epsilon}=\alpha^{-1}\boldsymbol{1}_{3},\;\boldsymbol{\mu}=\alpha\boldsymbol{1}_{3},\;\boldsymbol{\xi}=\left(\begin{matrix}0&\lambda&0\\-\lambda&0&0\\0&0&0\end{matrix}\right).\label{eq:example_material_2},
\end{align}
In the limit $\lambda\to1$ (\ref{eq:example_material_2}) again satisfies (\ref{eq:condition_example2}), and we predict that $\Psi_{+}'$ becomes an analytic function of $\mathcal{Z}=x+{\rm i}y$.  Similarly, for $\lambda\to-1$, $\Psi_{+}'$ becomes a function of $\mathcal{Z}^{\star}$.  This analyticity leads to unidirectional interface states, and we now consider the problem of waves trapped at the interface of two media where $\lambda=\pm1$.
\begin{figure}[h!]
    \centering
    \includegraphics[width=\textwidth]{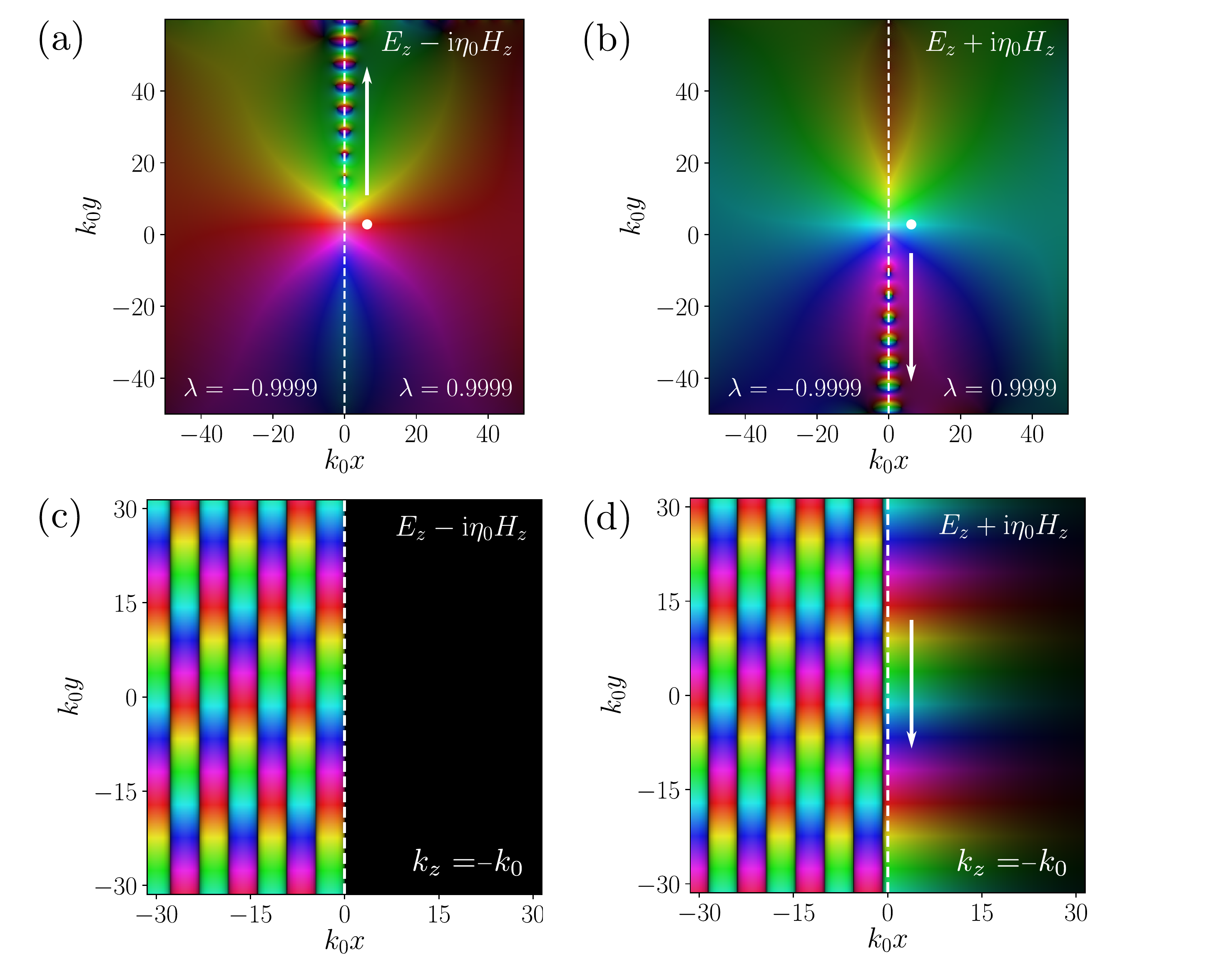}
    \caption{Panels (a) and (b) show a full wave simulation~\cite{comsol} of the field emitted from a magnetic line current within a bi--anisotropic medium with the constitutive relations (\ref{eq:example_material_2}).  Here $\alpha=1$ and $\lambda$ takes the indicated values on either side of $x=0$.  The saturation and colour indicate amplitude and phase as in Fig.~\ref{fig:scattering}.  Decomposing the field into the polarizations $E_{z}\pm{\rm i}H_z$ we see that each is constrained to propagate in only one direction away from the source.  See appendix~\ref{ap:comsol} for the modifications to COMSOL necessary to simulate bi--anisotropic materials.  Panels (c) and (d) plot the field (\ref{eq:electric_field_tir}--\ref{eq:magnetic_field_tir}) due to total internal reflection within a material with $\epsilon=1.5$, $k_z=-k_0$, and $k_y=0.2 k_0$.  In the two panels we resolve the field into the polarizations indicated, showing that only the field $E_{z}+{\rm i}\eta_0 H_z$ can propagate outside the dielectric when $k_y<0$.}
    \label{fig:edge_state_example}
\end{figure}
\par
For the parameters (\ref{eq:example_material_2}) Maxwell's equations take a very similar form to our previous example (\ref{eq:in-plane-fields})
\begin{align}
    \boldsymbol{\nabla}\Psi'_{+}\times\hat{\boldsymbol{z}}-{\rm i}\lambda\boldsymbol{\nabla}\Psi'_{+}&={\rm i}k_0\left(1-\lambda^{2}\right)\boldsymbol{\Phi}_{-}'\nonumber\\
    \boldsymbol{\nabla}\times\boldsymbol{\Phi}_{-}'&=-{\rm i}k_0\Psi'_{+}\label{eq:example2_fields}
\end{align}
where $\boldsymbol{\Phi}_{-}'=\alpha\eta_0\boldsymbol{H}_{\parallel}-{\rm i}\boldsymbol{E}_{\parallel}$.  Taking the curl of the first of (\ref{eq:example2_fields}) and applying the second equation shows that $\Psi_{+}'$ satisfies the same Helmholtz equation (\ref{eq:CAN_helmholtz}) as in the previous section.  Suppose now that we vary the bianisotropy in space according to $\lambda(x)=|\lambda|\,{\rm sign}(x)$ (i.e. we have an interface between two media that are the time reverse of one another).  From the continuity of the in--plane $\boldsymbol{E}$ and $\boldsymbol{H}$ fields across the interface, we know that at $x=0$ we must have continuity of both $\Psi_{+}'$ and $\hat{\boldsymbol{y}}\cdot\boldsymbol{\Phi}_{-}'$.  The second of these conditions applied to (\ref{eq:example2_fields}) implies
\begin{equation}
    -\frac{\partial\Psi'_{+}}{\partial x}+k_y|\lambda|\,{\rm sign}(x)\Psi'_{+}\qquad{\rm continuous\;at\;} x=0\label{eq:psi_p_continuity}
\end{equation}
where assumed propagation along the interface, $\partial_y\to{\rm i}k_y$.  Given that $\Psi_{+}'$ obeys the Helmholtz equation (\ref{eq:CAN_helmholtz}), which is insensitive to the sign of $\lambda(x)$, we must respectively have $k_x=\mp{\rm i}\kappa=\mp{\rm i}[k_y^{2}-k^2]^{1/2}$ on the two sides of the interface.  Applying this to the continuity condition (\ref{eq:psi_p_continuity}) then tells us the relation between the positive decay constant $\kappa$ and the propagation constant $k_y$
\begin{equation}
    \kappa=-k_y|\lambda|
\end{equation}
which can only be fulfilled for $k_y<0$, and implies a unidirectional interface state that satisfies the free space dispersion relation $k_y=-k_0$.    As predicted, when $|\lambda|=1$, $k_x=\pm{\rm i}k_y$, and the interface state becomes respectively an analytic function of $\mathcal{Z}^\star$ and $\mathcal{Z}$ on the two sides of the interface.  Note, although we used analyticity to predict the presence of this interface state, there is unidirectional propagation for a range of $\lambda$ values.  For any value of $\lambda>1$, propagation will be forbidden in the bulk of the material, and allowed only at the interface. Panels (a) and (b) of Fig.~\ref{fig:edge_state_example} show a full wave simulation of a magnetic line source next to the interface between the two media (\ref{eq:example_material_2}).  Decomposing the resulting field into the two polarizations $E_{z}\pm{\rm i}\eta_0 H_z$ shows that, as predicted each polarization is constrained to propagate in only direction along the interface.
\par
The material parameters given in Eq. (\ref{eq:example_material_2}) correspond to a time--irreversible bi--anisotropic medium, which may seem rather difficult to experimentally investigate.  However, in the case where $\boldsymbol{\xi}$ is homogeneous in space, this material can be obtained surprisingly simply.  It is equivalent to an isotropic material with $\epsilon=1/\mu$, through which a wave propagates with a fixed wave--vector $k_z=-\lambda k_0$ along the $z$--axis.  Such a fixed propagation is inherently time irreversible and mimics the above bi--anisotropy.  This can be seen through writing Maxwell's equations as e.g.
\begin{equation}
    \boldsymbol{\nabla}\times\boldsymbol{E}=\boldsymbol{\nabla}_{\parallel}\times\boldsymbol{E}+{\rm i}k_z\hat{\boldsymbol{z}}\times\boldsymbol{E}={\rm i}k_0\mu\eta_0\boldsymbol{H}    
\end{equation}
We can rearrange this equation so that the term proportional to $k_z$ appears as an effective bi--anisotropic response $\boldsymbol{\xi}^{\dagger}=-(k_z/k_0)\hat{\boldsymbol{z}}\times=\lambda\hat{\boldsymbol{z}}\times$, which is identical to the bi--anisotropy given in Eq. (\ref{eq:example_material_2}).  With this identification, we can assign an effective bi--anisotropy to any material, even free space.
\begin{figure}[h!]
    \centering
    \includegraphics[width=0.7\textwidth]{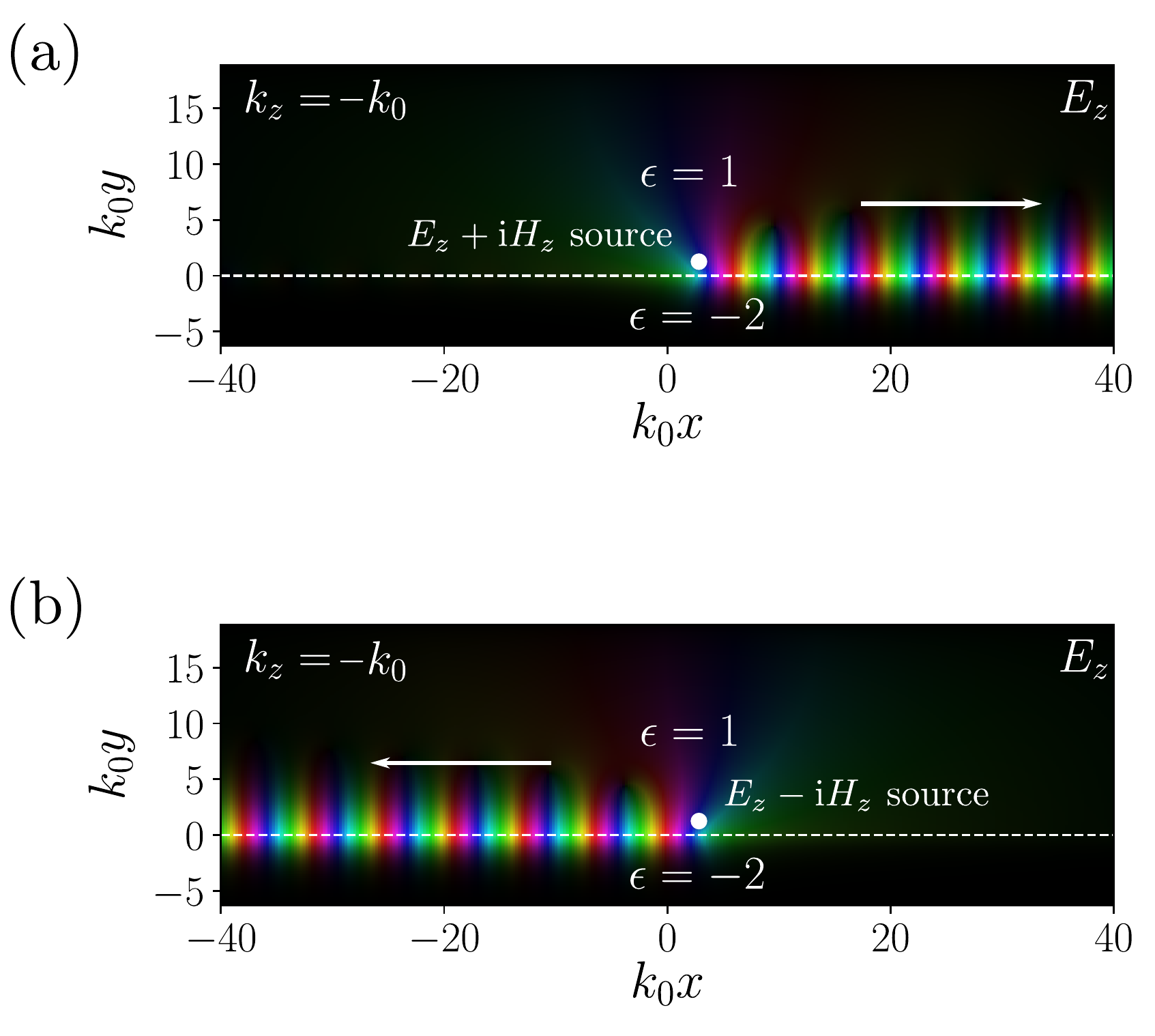}
    \caption{Full wave simulation~\cite{comsol} of a surface plasmon excited by a chiral antenna, where in contrast to e.g. Fig.~\ref{fig:edge_state_example} we plot the out of plane electric field $E_z$.  The chiral antenna excites only one of the polarizations $E_{z}\pm{\rm i}\eta_0 H_{z}$ and is modelled using an electric and magnetic line current aligned along the $z$--axis, positioned where the white dot is shown.  The amplitude of the magnetic line current is equal to that of the electric current times the free space impedance, and the phase is different by either (a) $+\pi/2$, or (b) $-\pi/2$.  Along the axis of the antenna the current has a propagation vector $k_z=-k_0$, so that propagation in the $x$--$y$ plane becomes equivalent to our CAN medium condition (\ref{eq:analyticity}).
    Saturation and colour indicate amplitude and phase as in Fig.~\ref{fig:scattering}}
    \label{fig:edge_state_example_2}
\end{figure}
\par
Rather surprisingly, using this  equivalence between bi--anisotropy and out of plane propagation in free space, the above unidirectional edge state is can be found in any evanescent wave.  In this particular case of our zero index condition, the unidirectional propagation is equivalent to the spin--momentum locking of evanescent waves recently identified by Bliokh and coworkers~\cite{bliokh2015}.
\par
The simplest case is total internal reflection.  We have a semi--infinite dielectric medium in the region $x<0$ with $\epsilon=n^{2}>1$.  A TE polarized wave is incident from inside the medium onto the interface at $x=0$.  The electric field is given by
\begin{equation}
    \boldsymbol{E}=\boldsymbol{e}_{1}\,{\rm e}^{{\rm i}\boldsymbol{k}_{\parallel}\cdot\boldsymbol{x}}\begin{cases}
    {\rm e}^{{\rm i}k_x x}+r\,{\rm e}^{-{\rm i}k_x x}&x<0\\
    t\,{\rm e}^{{\rm i}k_x'x}&x>0
    \end{cases}\label{eq:electric_field_tir}
\end{equation}
where $\boldsymbol{e}_{1}=\hat{\boldsymbol{x}}\times\hat{\boldsymbol{k}}_{\parallel}$, $\boldsymbol{k}_{\parallel}=k_y\hat{\boldsymbol{y}}-\lambda k_0\hat{\boldsymbol{z}}$, $k_x=\sqrt{\epsilon k_0^2-k_\parallel^2}$ and $k_x'=\sqrt{k_0^2-k_\parallel^2}$, and the reflection and transmission coefficients are
\begin{equation}
    r=\frac{k_x-k_x'}{k_x+k_x'},\;\;t=\frac{2k_x}{k_x+k_x'}
\end{equation}
The magnetic field corresponding to (\ref{eq:electric_field_tir}) can be calculated using the third Maxwell equation, $-{\rm i}k_0^{-1}\boldsymbol{\nabla}\times\boldsymbol{E}=\eta_0\boldsymbol{H}$, and is given by
\begin{equation}
    \eta_0\boldsymbol{H}={\rm e}^{{\rm i}\boldsymbol{k}_{\parallel}\cdot\boldsymbol{x}}\begin{cases}
    \boldsymbol{e}_{2}^{(+)}{\rm e}^{{\rm i}k_x x}+\boldsymbol{e}_{2}^{(-)}r\,{\rm e}^{-{\rm i}k_x x}&x<0\\
    \boldsymbol{e}_{2}^{'(+)}t\,{\rm e}^{{\rm i}k_x'x}&x>0
    \end{cases}\label{eq:magnetic_field_tir}
\end{equation}
where $\boldsymbol{e}_{2}^{(\pm)}=k_0^{-1}(k_\parallel\hat{\boldsymbol{x}}\mp k_x\hat{\boldsymbol{k}}_{\parallel})$, and $\boldsymbol{e}_{2}^{'(\pm)}=k_0^{-1}(k_\parallel\hat{\boldsymbol{x}}\mp k_x'\hat{\boldsymbol{k}}_{\parallel})$. \par
As discussed above, the region of free space above the surface ($x>0$) is equivalent to the bi--anisotropic medium (\ref{eq:example_material_2}) for the special case of $\alpha=1$.  Using Eqns. (\ref{eq:electric_field_tir}--\ref{eq:magnetic_field_tir}) to calculate the combination of out of plane field components $\Psi_{+}'=E_{z}+{\rm i}\eta_0 H_z$ we find
\begin{equation}
    \Psi'_{+}=E_{z}+{\rm i}\eta_0 H_z=\frac{t}{k_\parallel}\left(k_y+{\rm i} k_x' \lambda\right){\rm e}^{{\rm i}(k_x'x+k_y y)}{\rm e}^{-{\rm i}\lambda k_0 z}\label{eq:psi_plus_prime}
\end{equation}
In the limit $\lambda\to+1$, the wave inside the dielectric propagates down the $z$ axis with a wavenumber equal to the free space wavenumber $k_0$.  Therefore for all values of $k_y$ the wave will be totally internally reflected and exponentially decays in the region of free space outside.  However, because $\Psi_{+}'$  becomes an analytic of $\mathcal{Z}$, Eq. (\ref{eq:psi_plus_prime}) is only non--zero when $k_y<0$
\begin{equation}
    \lim_{\lambda\to1}\Psi_{+}'=-\frac{2|k_y|t}{k_{\parallel}}
    \begin{cases}
    0&k_y>0\\
    {\rm e}^{-|k_y|(x+{\rm i}\,y)}&k_y<0
    \end{cases}
\end{equation}
Therefore for total internal reflection where the internal propagation angle is such that $k_z=-k_0$, the field component $\Psi_{+}$ outside becomes an analytic function of $\mathcal{Z}=x+{\rm i}y$ (and similarly $\Psi'_{-}$ becomes an analytic function of $\mathcal{Z}^{\star}$).  We can therefore see that if we work in the polarization basis $E_{z}\pm{\rm i}\eta_0 Hz$, the two polarizations are constrained to propagate in only one direction along the surface.  Panels (c) and (d) in Fig.~\ref{fig:edge_state_example} illustrate this phenomenon.  Here the wave is incident from inside the dielectric with $k_z=-k_0$ and $k_y=-0.2 k_0$.  Plotting the two polarizations $\Psi'_{\pm}$, we see that only $\Psi'_{+}$ is non zero in the region outside the dielectric.  This is independent of the form of the transmission coefficient, and would be true for any evanescent wave, as pointed out in~\cite{bliokh2015}.
\par
To illustrate the generality of this finding, in Fig.~\ref{fig:edge_state_example_2} we show the out of plane electric field $E_z$ obtained from a full wave simulation of a chiral source in free space next to a metal $\epsilon=-2$.  The source is imagined to be a chiral antenna extended along the $z$--axis, with propagation vector $k_z=-k_0$ along the axis of the antenna.  The source is modelled using a combined electric current $j_z=J{\rm e}^{-{\rm i}k_0 z}$ and magnetic current $m_z=\pm{\rm i}\eta_0 J{\rm e}^{-{\rm i}k_0 z}$ at the point indicated with the white dot.  Such a source (which incidentally could possibly be realised using a variant of the electromagnetic line mode investigated in~\cite{horsley2014,bisharat2017}) excites only the $\Psi_{\pm}'$ polarization, which as illustrated only propagates in one direction along the surface.  
%
%
\section{Summary and conclusions}
\par
In this work we investigated the general conditions for a linear electromagnetic material to exhibit zero refractive index in a specified direction of propagation.  It was found that the $6\times 6$ material tensor $\chi$ defined in Eq. (\ref{eq:chi_def}) must have an eigenvector with eigenvalue zero.  We illustrated the generality of this condition through enforcing it (via Eq.~\ref{eq:projection}) on several examples, including a randomly generated set of material parameters.  
\par
The purpose of finding this general zero index constraint was to formulate a condition for a material to support waves that propagate in only one direction.  In a previous publication~\cite{Horsley2017UnidirectionalAxes} it was found that in the special case of gyrotropic media, unidirectional propagation can be associated with a zero of the refractive index in a complex direction.  Here we have found the generalization of this result, and it is now clear that there are an infinite family of homogeneous materials that support such uni--directional propagation, which we have called complex axis nihility (CAN) media.  We have given two specific bi--anisotropic examples that have known experimental realizations.  These results tally with existing findings such as that of~\cite{bliokh2015}, and in the simplest case we have shown that in the polarization basis $E_{z}\pm{\rm i}\eta_0 H_z$, any evanescent wave (such as e.g. a surface plasmon, or a totally internally reflected wave) can be understood as a unidirectional edge mode.  One can understand this in terms of either reciprocal space topology and spin--momentum locking (as in~\cite{bliokh2015}), or as we have done here, in terms of an effective bi--anisotropy due to out of plane propagation.
\par
Finally it worth noting that, although it remains to show their equivalence, applying our zero index condition is much simpler than topological methods.  Comparison with these methods shows that perhaps we can view the condition (\ref{eq:analyticity}) as a means to find the edge of a pass band that is separated by a band gap in parameter space from another pass band, with a uni--directional interface state joining the two.  This is certainly the case in the examples given here and in~\cite{Horsley2017UnidirectionalAxes}.  Note also that we are claiming no `topological robustness' of our modes to defects.  This is a difficult claim to make in any case, scatterers typically couple polarizations, and nearly every system investigated to date is only robust to some special class of scatterers; here is no exception.  However, even without such robustness, unidirectional propagation can still be very useful. For example, the plasmonic example shown here and in~\cite{bliokh2015} is not robust to e.g. imperfections in the surface, but illustrates the possibility to launch a surface wave in only one direction without using a phased array (which is necessarily a much larger device).
\acknowledgements
SARH acknowledges financial support from a Royal Society TATA University Research Fellowship (RPG-2016-186). MW acknowledges funding from an EPSRC vacation bursary.  SARH acknowledges useful conversations with W. L. Barnes and I. R. Hooper, as well as I. R. Hooper's numerical expertise.
%
%
\appendix
\noindent
\section{Material parameters\label{ap:random}}
In Figure~\ref{fig:random_example} we illustrated the application of our general formula (\ref{eq:chi_proj}) to an arbitrary bi--anisotropic material.  In this case the initial material tensor $\chi_{\rm i}$ was generated using a random number generator.  For reference we give the tensor here
\begin{equation}
\small
    \chi=\left(\begin{matrix}1.454&0.205-0.115{\rm i}&0.401-0.167{\rm i}&0.301+0.073{\rm i}&0.353+0.229{\rm i}&0.301+0.127{\rm i}\\
    0.205+0.115{\rm i}&1.591&0.108-0.235{\rm i}&0.296+0.186{\rm i}&0.234-0.161{\rm i}&0.162-0.094{\rm i}\\
    0.401+0.167{\rm i}&0.108+0.235{\rm i}&1.322&0.357+0.041{\rm i}&0.417-0.190{\rm i}&0.087-0.218{\rm i}\\
    0.301-0.073{\rm i}&0.296-0.186{\rm i}&0.357-0.041{\rm i}&1.534&0.481+0.101{\rm i}&0.647-0.118{\rm i}\\
    0.353-0.229{\rm i}&0.234+0.161{\rm i}&0.417+0.190{\rm i}&0.481-0.101{\rm i}&1.319&0.299-0.385{\rm i}\\
    0.301-0.127{\rm i}&0.162+0.094{\rm i}&0.087+0.218{\rm i}&0.647+0.118{\rm i}&0.299+0.385{\rm i}&1.143\end{matrix}\right).\label{eq:chi0_random}
\end{equation}
%
%
\section{Modifications to the COMSOL Multiphysics constitutive relations\label{ap:comsol}}
\par
In order to use COMSOL multiphysics to simulate propagation in materials with the bi--anisotropic constitutive relations (\ref{eq:constitutive_relations}) we needed to modify the equations in the radio frequency module.  Here we give details of these modifications.  First we define a set of variables as `global definitions'; {\verb xixx,xixy,... } and {\verb zexx,zexy,... }, as well as defining the shortened constant names {\verb e0 } and {\verb m0 } for {\verb epsilon0_const } and {\verb mu0_const } respectively.  Then the expression for the displacement field is modified to

\begin{Verbatim}[fontsize=\small]
Dx = e0*(emw.epsilonrxx*emw.Ex+emw.epsilonrxy*emw.Ey+emw.epsilonrxz*emw.Ez)
+(1/c_const)*(xixx*emw.Hx+xixy*emw.Hy+xixz*emw.Hz)

Dy = e0*(emw.epsilonryx*emw.Ex+emw.epsilonryy*emw.Ey+emw.epsilonryz*emw.Ez)
+(1/c_const)*(xiyx*emw.Hx+xiyy*emw.Hy+xiyz*emw.Hz)

Dz = epsilon0_const*(emw.epsilonrzx*emw.Ex+emw.epsilonrzy*emw.Ey+emw.epsilonrzz*emw.Ez)
+(1/c_const)*(xizx*emw.Hx+xizy*emw.Hy+xizz*emw.Hz)
\end{Verbatim}
the polarization field to
\begin{Verbatim}[fontsize=\small]
Px = e0*(emw.epsilonrxx*emw.Ex+emw.epsilonrxy*emw.Ey+emw.epsilonrxz*emw.Ez)
+(1/c_const)*(xixx*emw.Hx+xixy*emw.Hy+xixz*emw.Hz)- e0*emw.Ex

Py = e0*(emw.epsilonryx*emw.Ex+emw.epsilonryy*emw.Ey+emw.epsilonryz*emw.Ez)
+(1/c_const)*(xiyx*emw.Hx+xiyy*emw.Hy+xiyz*emw.Hz)-e0*emw.Ey

Pz = e0*(emw.epsilonrzx*emw.Ex+emw.epsilonrzy*emw.Ey+emw.epsilonrzz*emw.Ez)
+(1/c_const)*(xizx*emw.Hx+xizy*emw.Hy+xizz*emw.Hz)-e0*emw.Ez  
\end{Verbatim}
the $\boldsymbol{H}$ field to
\begin{Verbatim}[fontsize=\small]
Hx = (emw.murinvxx*(emw.Bx-(1/c_const)*(zexx*emw.Ex+zexy*emw.Ey+zexz*emw.Ez))
+emw.murinvxy*(emw.By-(1/c_const)*(zeyx*emw.Ex+zeyy*emw.Ey+zeyz*emw.Ez))
+emw.murinvxz*(emw.Bz-(1/c_const)*(zezx*emw.Ex+zezy*emw.Ey+zezz*emw.Ez)))/m0

Hy = (emw.murinvyx*(emw.Bx-(1/c_const)*(zexx*emw.Ex+zexy*emw.Ey+zexz*emw.Ez))
+emw.murinvyy*(emw.By-(1/c_const)*(zeyx*emw.Ex+zeyy*emw.Ey+zeyz*emw.Ez))
+emw.murinvyz*(emw.Bz-(1/c_const)*(zezx*emw.Ex+zezy*emw.Ey+zezz*emw.Ez)))/m0

Hz = (emw.murinvzx*(emw.Bx-(1/c_const)*(zexx*emw.Ex+zexy*emw.Ey+zexz*emw.Ez))
+emw.murinvzy*(emw.By-(1/c_const)*(zeyx*emw.Ex+zeyy*emw.Ey+zeyz*emw.Ez))
+emw.murinvzz*(emw.Bz-(1/c_const)*(zezx*emw.Ex+zezy*emw.Ey+zezz*emw.Ez)))/m0
\end{Verbatim}
and the time derivative of the $\boldsymbol{H}$ field to
\begin{Verbatim}[fontsize=\small]
dHdtx = emw.iomega*emw.Hx

dHdty = emw.iomega*emw.Hy

dHdtz = emw.iomega*emw.Hz.
\end{Verbatim}
%
%
%
\bibliography{refs}
\bibliographystyle{unsrt}
\end{document}